# Sub-femtosecond stabilization of multicore fiber for high-fidelity quantum networking at 100% duty cycle

**Takuma Nakamura[1,2]\*, Nazanin Hoghooghi[1], Nicolas Fontaine[3], Tetsuya Hayashi[4], Takuji Nagashima[4], Nicholas V. Nardelli[1], Dileep V. Reddy[5], Martin J. Stevens[5], Tara Fortier[1], Lynden K. Shalm[5,6], and Franklyn Quinlan[1,6]■**

[1]Time and Frequency Division, National Institute of Standards and Technology, 325 Broadway, Boulder, CO 80305, USA
[2]Department of Physics, University of Colorado Boulder, 440 UCB Boulder, CO, 80309, USA
[3]Nokia Bell Labs, Murray Hill, NJ, USA
[4]Sumitomo Electric Industries, Ltd., 1, Taya-cho, Sakae-ku, Yokohama, Kanagawa, Japan
[5]Applied Physics Division, National Institute of Standards and Technology, 325 Broadway, Boulder, CO 80305, USA
[6]Electrical, Computer and Energy Engineering, University of Colorado, Boulder, Colorado 80309, USA
\*takuma.nakamura@nist.gov; ■fquinlan@nist.gov

**Abstract**

Originally envisioned as a solution for the capacity crunch in telecommunications networks, multicore fibers (MCF) are contributing to scientific fields beyond telecom, such as sensing and metrology. Confined within the same cladding, the cores of MCF have a high degree of noise correlation which can be harnessed for a variety of applications. Here, we investigate MCF as a solution to the challenging problem of quantum and classical light co-existence in quantum networks by operating the quantum and stabilization light in separate but highly correlated cores of a 7-core MCF. Over 40 km of spooled fiber, we achieved 100 attosecond integrated jitter on one core by using phase information derived from another core. This allows for 100% duty cycle on a quantum channel while maintaining a low spurious photon rate from crosstalk between stabilization and quantum channels. With cycle-slip-free stabilization over 6 hours, frequency detuning between designated stabilization and quantum channels, and an additional 40 dB rejection of noise photons provided by the low optical crosstalk between cores, we achieved a Raman scattering-induced spurious photon rate of only 0.01 photons/s in 100 GHz bandwidth. Our results with MCF are a promising approach to ultra-stable quantum networks with 100% duty cycle on the quantum channel.

**Introduction**

Co-existence of quantum and classical light in optical fibers is a challenge for real-world implementations of quantum networks. Achieving better than fs-level synchronization needed for many quantum networking protocols [1–3] requires classical light (cw or pulsed) to co-propagate with the quantum light and be used for canceling the phase noise added by the fiber. Importantly, to reach this level of synchronization both the classical and quantum light must share the same fiber, since fiber pairs, even when bundled together, do not share the requisite level of timing noise correlation [4,5].

A challenge arises in co-existence of quantum and classical light (data and/or synchronization light) over fiber due to the strong classical light directly or indirectly coupling into the quantum channel. Direct coupling can be mitigated by operating the quantum channel at a different wavelength within a different International Telecommunication Union (ITU) channel from the bright classical light, but comes at the cost of added loss from the optical filters on the quantum channel. Indirect coupling, mainly through spontaneous Raman scattering [6], is more difficult to eliminate. Any Raman photons (noise photons) generated by the classical light at the wavelength of the quantum signal cause spurious photon counts and reduced fidelity. Strategies for co-existence have mainly consisted of wavelength division multiplexing (WDM), time division multiplexing (TDM) or a combination of both. In WDM schemes, the wavelength of the quantum light could be selected based on the lowest point in the Raman gain spectrum in the fiber. In this way, the number of generated Raman photons at the quantum signal wavelength is reduced. Operating with quantum light <1300 nm has been demonstrated over single mode fiber [7–9]. However, these wavelengths do not correspond to the lowest loss in fiber which is an important consideration in quantum networks. Additionally, it is advantageous to further reduce the level of Raman-scattered photons in the quantum channel. Time multiplexing is an effective method for enabling co-existence over fiber, but this method always comes at the price of reduced throughput or duty cycle of the quantum channel [10].

Multicore fibers (MCF), where several cores are confined within the same cladding [11], offer an elegant solution to the co-existence problem without reducing the duty cycle or increasing loss. The cores within a MCF with uncoupled-core design guide light in a single spatial mode with loss similar to standard single mode fiber (SMF), while the proximity of the cores inside the cladding translates to a high degree of noise correlation between them [5]. Figure 1 shows the concept of co-existence in a MCF fiber link that takes advantage of this noise correlation. By using separate cores of an MCF for quantum and classical signals, the two channels are physically separated. At the same time, the high degree of correlation of environmentally induced instabilities between cores allows for extremely low relative timing jitter, such that the quantum channel is effectively synchronized with stabilization light traveling through a separate core (classical channel). The core-to-core proximity will induce optical crosstalk, the level of which depends on parameters such as the design (e.g., with or without trenches around the core), number of cores, wavelength of the light and whether the fiber is spooled or deployed. This crosstalk will introduce spurious photons from the stabilization light into the quantum channel. However, crosstalk as low as -40 dB is readily achievable, and can even reach the -80 dB level [11]. When combined with frequency detuning of the stabilization light, MCF will significantly reduce the number of Raman-induced noise photons coupling into the quantum channel. Quantum key distribution (QKD) over MCF with classical and quantum light in different cores has been demonstrated previously [12–14]. Although these studies demonstrated mitigation of Raman photon induced noise through WDM or TDM or both, no quantitative measurement of the Raman-scattered photon number has been shown, leaving the exact contribution of Raman scattering uncharacterized. Furthermore, these QKD demonstrations are not sensitive to the timing jitter from the fiber link, and sub-femtosecond jitter has yet to be demonstrated in MCF studies targeting quantum networking applications.

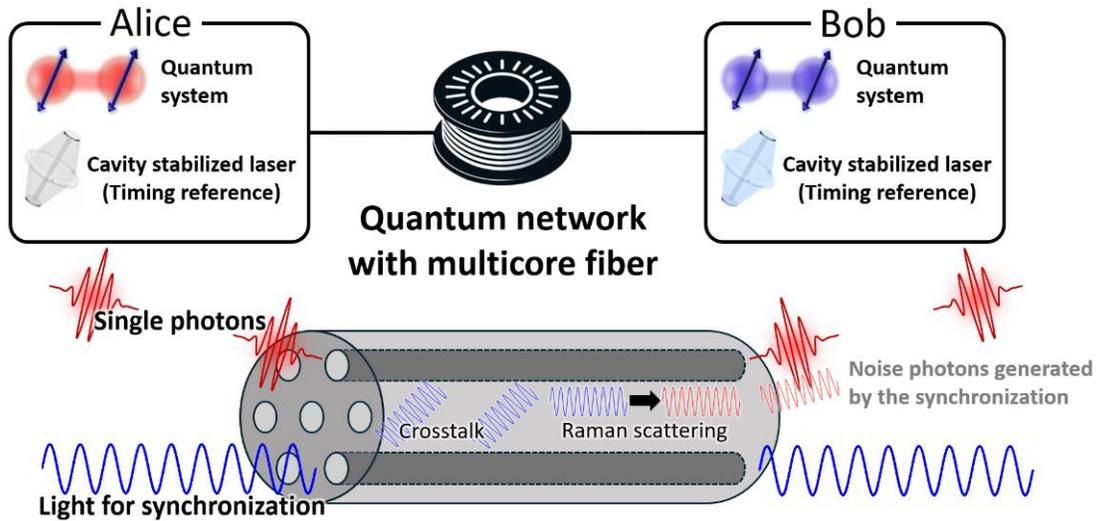

Figure 1. MCF as a solution for classical and quantum light co-existence in quantum networks. Quantum and classical (synchronization) light are sent over separate but highly noise correlated cores of a MCF, allowing for high network throughput while providing enough SNR signal for synchronization. The number of Raman scattering photons coupled into the quantum channel is significantly reduced compared to SMF.

In this work, we experimentally investigate how MCF can enable the co-existence of quantum and classic light using 40 km of a weakly coupled 7-core fiber. We take advantage of the high degree of noise correlation between cores to effectively stabilize the optical path length at 1542 nm in one core ("quantum channel") by using light for active fiber noise cancellation at 1550 nm in a separate core ("stabilization channel"). Using this method, the fiber-induced integrated timing jitter on the quantum channel is only 400 attoseconds (1 MHz to 0.5 Hz) across the 40 km length of fiber, reducing to 100 attoseconds when removing the noise contribution of the source laser. Furthermore, we measured the number of spontaneous Raman scattering photons coupled into the quantum channel using a superconducting nanowire single photon detector (SNSPD). By reducing the launched stabilization light power required for cycle-slip-free phase locking to only ~400 pW (corresponding to 300 fW at the detector), the scattered light into the quantum channel is estimated to be less than 0.01 photons/s in 100 GHz optical bandwidth. This is ~4 orders of magnitude lower than that achieved when a single core is used for wavelength-multiplexed stabilization and quantum channels. The demonstrated level of spurious photons is comparable to the dark count rate of state-of-the-art single photon detectors, providing a path towards high-fidelity, ultra-stable quantum networks with 100% duty cycle.

**Experiments**

*1) Phase noise and long-term stability analysis*

A schematic of our experimental setup for studying co-existence in MCF is shown in Fig. 2(a). We used a 40 km-long spool of MCF with 7 independent cores confined within a 160 μm diameter cladding [15]. Each end of the fiber is spliced to a fan-in/fan-out (FI/FO) device which breaks out the 7 cores to individual single mode fibers. The total loss of the fiber, including the loss FI/FO devices, at 1550 nm was measured to be ~10 dB. Assuming the loss of the MCF is consistent with the ~0.2 dB/km of standard SMF fiber (8 dB loss in 40 km), the total link loss is as expected. The two cores we chose as quantum and stabilization channels are designated core-2 and core-7, respectively. This pair was chosen based on the highest degree of noise correlation between them.

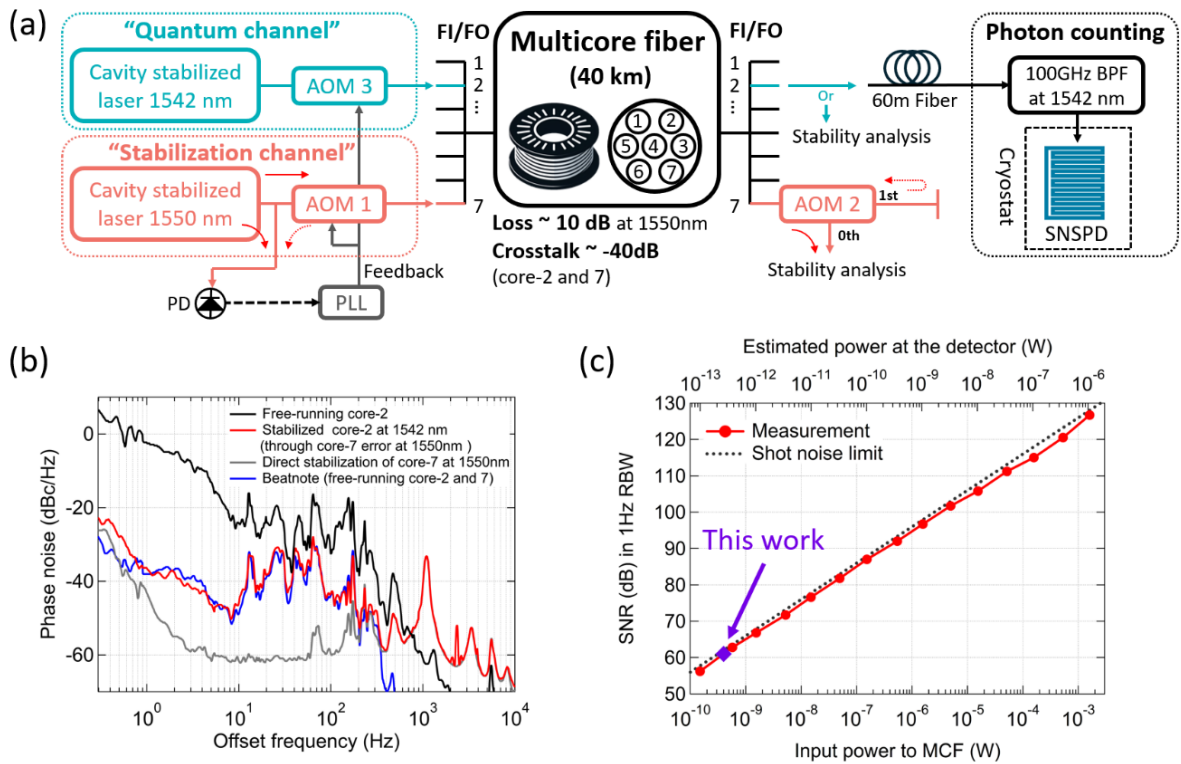

Figure 2. (a) Schematic of the experimental setup of the MCF fiber link. (b) Phase noise spectra of the free-running core-2 (black), stabilized core-2 through stabilizing core-7 (red), stabilized core-7 (gray) and beat note between the free-running core-2 and core-7 (blue). For red and gray traces, we use only 400 fW of optical power, measured at the detector, for stabilization. (c) The measured relationship between the SNR of the beat note for the phase lock and the input power to the MCF (red). The shot noise limited SNR (dotted line) calculated from the estimated power at the detector for the beat detection (top axis).

We used cavity-stabilized laser light at 1542 nm [16,17] to represent the quantum channel in core-2. This 8 nm (~1 THz) detuning is to reduce direct coupling of the stabilization light to the quantum channel, allow for the use of single photons with broader bandwidth (~1 nm) and retain a high degree of noise correlation between channels. Cavity-stabilized laser light is used to ensure the noise of the laser is below that of the fiber for most noise offset frequencies of interest. In these experiments, the noise below 1 kHz offset frequencies is dominated by the fiber noise. Crosstalk between core-2 and core-7, including the FI/FO devices, was measured to be ~-40 dB. The effect of crosstalk on the coupling or leakage of light into the quantum channel is discussed in detail below.

We employed active Doppler cancellation for fiber stabilization [18]. The light for each channel was frequency-shifted with an acousto-optic modulator (AOM), labeled as AOM1 and AOM3 in Fig. 2(a), before launching into its respective core. We note that these AOMs could be replaced with a fiber stretcher to minimize the loss. Another AOM (AOM2) was placed at the output end of the stabilization channel to distinguish the light that travelled the entire fiber link from stray reflections along the fiber path, such as at splicing points or fiber mating sleeves. The zeroth-order output from AOM2 was used for stability analysis of core-7 by taking a beat note with the source laser. The first order was reflected back along the 40 km fiber path by a retroreflector. The reflected light passes through AOM1 again and interferes with the original laser light to create a heterodyne beat used for stabilization. The frequency of this heterodyne beat note is the sum of the doubled frequencies (due to double-passing through the AOMs) driving AOM1 (100 MHz) and AOM2 (55 MHz), onto which is added the frequency noise of the fiber link. The 310 MHz signal was phase locked to a reference synthesizer by applying correction to the driving source of AOM1. The high degree of noise correlation between core-2 and core-7 allows us to apply the same feedback signal to the AOM3 (quantum channel) through the common driving source at 100 MHz.

Fiber-induced phase fluctuations will be slightly different for different optical frequencies, so that the phase correction derived at one wavelength should be scaled before applying it to another wavelength for optimal fiber noise suppression [19]. This scaling was not applied for the 8 nm wavelength difference used in this experiment, limiting the amount of phase noise suppression to ~46 dB. As shown by the phase noise spectra of Fig. 2(b) the finite noise correlation between cores limits the core-to-core noise suppression to above this level. However, longer-term measurements approach this frequency-difference limit.

Due to the finite optical crosstalk between cores, we explored the minimum light power required for robust stabilization of the fiber path to reduce the number of spurious photons that couple to the quantum channel. Even for stable phase locks that are maintained over long time periods, there remains a finite probability of a "cycle slip", meaning the feedback control circuit momentarily loses track of the beatnote phase before regaining lock. In addition to momentarily increasing the path length jitter, slipping one or more optical cycles can result in an effective path length shift along the fiber as the feedback circuit reacquires lock on a different interference fringe. The probability of a cycle slip is strongly influenced by the signal-to-noise ratio (SNR) of the phase-locked signal when the light power is low [20]. With shot noise-limited detection, the influence the broadband white noise floor has on the lock can be mitigated by filtering the signal at the input of the feedback servo. However, reducing the signal bandwidth too much will effectively reduce the feedback gain bandwidth, thereby reducing the suppression of the fiber noise. Thus, a balance must be struck that depends on the free-running fiber noise, the optical power used for stabilization, the allowable spurious photon count in the quantum channel, and the system tolerance to the occasional cycle slip.

The relationship between the obtained SNR of the beat note for phase locking and input power to core-7 is shown in Fig. 2(c). Since we create a heterodyne beat with a strong local oscillator (source laser light), we achieve shot noise-limited SNR even with sub pW power on the reflected light at the detector. The SNR limit expected from shot noise is shown with the dotted line. The purple dot indicates an input power of 400 pW (300 fW on the detector used in the feedback loop) resulting in SNR of ~60 dB in 1 Hz resolution bandwidth (60 dBc/Hz). This was the lowest SNR that allowed us to achieve cycle-slip-free phase locking for continuous measurements up to several hours in duration.

Whereas this low power for stabilization will help reduce spurious photon counts in the quantum channel, high SNR is needed to accurately evaluate the residual phase noise and instability of the fiber. Attenuating the light returning from the remote end allows us to keep the power on the feedback detector low, and thereby maintain the same locking conditions, while providing plenty of light for noise and stability evaluations at the fiber link output. The return light was reduced by decreasing the efficiency of AOM2 through lowering the RF power of the drive signal. In this way, with mW-level power launched into core-7 we maintained enough power in the AOM's zeroth order for analysis with high SNR while the SNR on the feedback detector was reduced to only 60 dBc/Hz. To measure the relative stability of core-2, we launched more than 100 μW of 1542 nm light into core-2 and measured the added phase noise relative to the source 1542 nm laser.

The stabilization results are shown in Fig. 2(b). The black trace shows the free-running phase noise power spectral density of core-2. Core-2 and core-7 have very similar free-running noise spectra, such that the black trace is representative of core-7 as well. When core-7 is stabilized, its phase noise is suppressed by ~40 dB, with phase noise at its output shown in the gray trace. The phase noise of the stabilized core-7 reaches a minimum of ~60 dBc/Hz, matching the expected level from the SNR on the feedback detector. The prominent peak around 1.2 kHz is attributed to the servo feedback (servo bump). The offset frequency of this peak is determined by $\frac{1}{4\tau}$ where $\tau$ is the one-way delay of light over the fiber. For a 40 km link $\tau$ is ~200 μs corresponding to a peak at 1.25 kHz. Note that the amplitude of the peak is enhanced due to the poor SNR which requires tighter lock parameters to prevent cycle slips. This servo bump is suppressed by 10 dB when more optical power was used for the stabilization. Also, for shorter fiber lengths, the servo bump is shifted to higher frequencies and lower amplitude, which significantly reduces the total timing jitter. Therefore, care must be taken when comparing fiber link stabilization results with different fiber lengths.

With the correction signal derived from core-7 also applied to core-2, the residual phase noise of core-2 was measured, shown in the red trace of Fig. 2(b). Although its phase noise does not reach the stabilized channel level (gray trace), we achieve >20 dB noise suppression for offset frequencies below 10 Hz. The integrated jitter/phase noise is 400 attoseconds / 500 mrad, integrated from 1 MHz from 0.5 Hz. Importantly, noise from the source laser adds 100 attoseconds to the integrated timing jitter for offset frequencies above 10 kHz due to the delayed self-heterodyne effect [21], and the contribution of the servo bump at 1.2 kHz determined by fiber length was about 200 attoseconds. Thus, when the noise integration upper frequency bound is limited from 1 kHz to 0.5 Hz, the integrated jitter reduces to 100 attoseconds.

The difference between the phase noise of stabilized core-2 and core-7 in Fig. 2 (b) stems from the limited correlation between the cores. To verify this, the core-to-core correlation was measured directly. In this measurement, the same 1550 nm laser was split and launched to both cores simultaneously. At the end of the 40 km fiber, the relative phase between the two cores was measured by taking the heterodyne beat note between the outputs. Note that this measurement was without active fiber stabilization. For perfectly correlated cores, no residual noise should be observed. The relative phase noise between core-2 and core-7 is shown in the blue trace of Fig. 2 (b). For offset frequencies below ~1 kHz, this closely matches the noise of core-2 when the correction derived from core-7 is applied (red trace), verifying that the stability of core-2 is limited by the noise correlation between core-2 and core-7.

In addition to the phase noise power spectrum, long-term phase fluctuations over 22000 s (6.1 hours) were measured, with results shown in Fig. 3. The black trace in Fig. 3(a) shows the phase evolution of the 1542 nm light in core-2 of the MCF without fiber path stabilization. Over the course of the measurement, the phase fluctuations exceed 14000 radians, corresponding to a peak-to-peak timing error of ~10 ps. When stabilized using the correction signal derived from core-7, the 1542 nm phase fluctuations in core-2 are dramatically reduced, as shown in the red trace. A zoomed-in plot of core-2 fluctuations while stabilized via core-7, shown by the red trace in Fig. 3 (b), reveals peak-to-peak timing errors are reduced to ~25 fs. For comparison, phase fluctuations of core-7 when stabilized are also shown. Excess phase fluctuations on core-2 above those of core-7 are attributed to two sources. First, non-common fiber paths, in particular the FI/FOs, will exhibit a relative phase drift over the hours-long measurement duration. This contribution was estimated by bypassing the MCF, with resulting long-term phase variations consistent with the data in Fig. 3(b). Second is the frequency-difference limit, given by the fact that we measure the phase fluctuations at 1550 nm and apply the correction to 1542 nm.

A statistical measure of the long-term phase stability is given in Fig. 3(c) in terms of time deviation (TDEV) [22]. With stabilization on core-7, the TDEV for core-2 remains below 1 fs for averaging times up to several hundred seconds, though again we see noise on core-2 in excess of core-7. For averaging times from 0.1 s to 1 s, the excess noise is due to the limited core-to-core correlation seen in the phase noise of Fig. 2 (b).

For averaging times 100 s and above, core-2 is limited by non-common fiber paths and the frequency-difference limit.

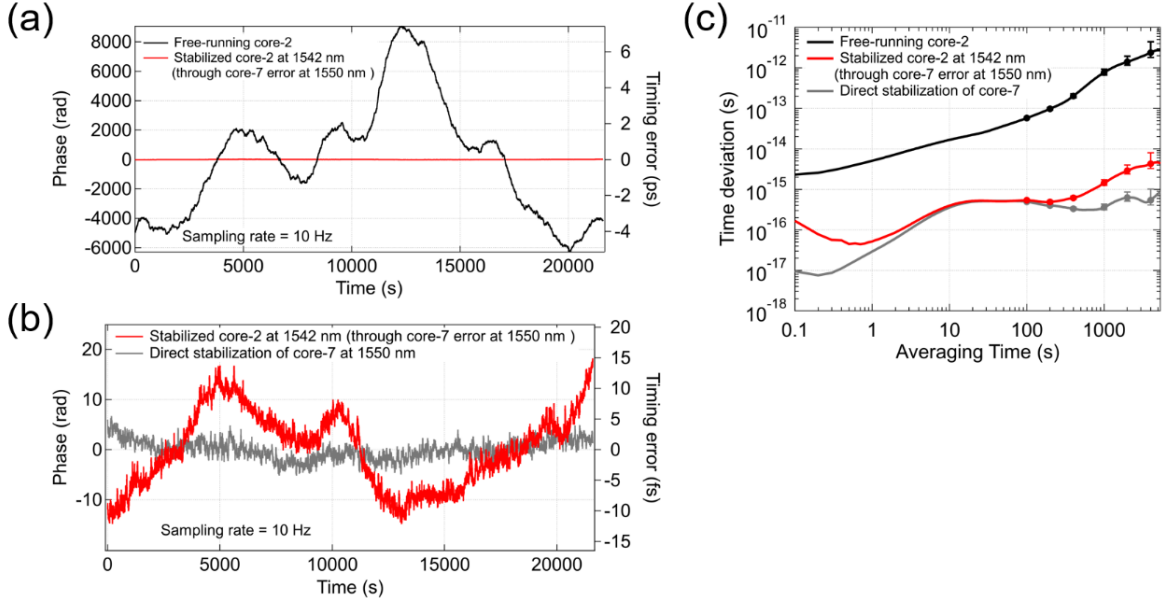

Figure 3. Long-term measurements of the fiber link stability. (a) Phase and timing error of the free-running core-2 (black) and core-2 with stabilization through core-7 error (red). (b) Zoomed in traces of the red in (a) and the direct stabilization of core-7 (gray). (c) Time deviation of the same data set. Free-running core-2 (black), core-2 with stabilization through core-7 error (red) and, the direct stabilization of core-7 (gray).

2) *Counting Raman Photons in the Quantum Channel*

In this section, we investigate the number of noise photons per second, $\Phi_q$, coupled into the quantum channel as we change the launched light power in the stabilization channel. We can estimate $\Phi_q$ using :

$$\Phi_q = \alpha_{fiber} \times XT \times P_{in}/h\upsilon \times S_{Raman} \times BW \qquad (Eq. 1)$$

where $\alpha_{fiber}$ is the loss of fiber, XT is the core-to-core crosstalk of the MCF, $P_{in}$ is the launched stabilization light power, $h\upsilon$ is photon energy, $S_{Raman}$ is an effective Raman scattering coefficient, and BW is an effective optical bandwidth of quantum channel.

For a quantitative estimate, we assume quantum and stabilization light are at 1542 nm and 1550 nm, respectively, as above. Normalized optical power spectral densities of the 1550 nm light signal measured with an optical spectrum analyzer both at the input (black trace) and the output (red trace) of core-7 are shown in Fig. 4(a). Spontaneous Raman scattering from the 1550 nm stabilization signal generates noise photons over a broad spectrum. At the output, two cascaded DWDMs (dense wavelength division multiplexers) centered at 1542 nm with 100 GHz bandwidth selected the light power at the quantum channel wavelength, shown in the blue trace. This filtered spectrum represents the background Raman photon level when using a single core for both classical and quantum channels. After accounting for the spectrum analyzer's measurement resolution bandwidth, the noise is ~65 dB/nm below the laser peak. We also measured -40 dB crosstalk between core-2 and core-7, 10 dB loss through the fiber and minimum cycle-slip free stabilization input power to the MCF of 400 pW. From these measurements and using Eq. (1), we estimate the number of noise photons per second, $\Phi_q$, that would couple to a quantum channel in core-2 integrated over a 100 GHz bandwidth near 1542 nm at the output of the 40 km link as $\sim 8 \times 10^{-3}$ photons/s.

To verify this estimate, we used a SNSPD [23] to count the number of spurious Raman photons at 1542 nm coupled from core-7 to core-2. The result of this measurement as a function of input power to core-7 (stabilization channel) is shown in Fig. 5. Note that the attenuation of the stabilization light happens at the input of the MCF, unlike the previous phase noise measurements where AOM3 was used to attenuate the return light. As the SNSPD was in a separate lab, a 60 m long delivery fiber was used to connect the output of the MCF to the SNSPD input. Knowing the loss of the delivery fiber and the insertion loss of the two 1542 nm DWDMs, we calculated the effective number of photons at the output of MCF from the measured photon numbers. We repeated the measurement using only core-7 and measuring the number of Raman photons in the 1542 nm channel at the output of core-7. This creates a useful benchmark to compare the MCF performance when using multiple cores with one-core performance, a good approximation to SMF. The measurement floor is limited by the background count rate of the SNSPD, which is likely dominated by room-temperature thermal photons coupling into the fibers. Both core-2 and core-7 measurements are extrapolated by subtracting the background count rate and assuming the spontaneous Raman scattering is linear with the input power.

The extrapolated core-2 measurement is about 40 dB lower than that of core-7, as expected from measurements of the core-to-core crosstalk. Importantly, the crosstalk is believed to be limited by the FI/FOs and not the MCF itself, such that this result can be further improved with higher performance FI/FOs [24]. The purple dotted line in Fig. 4 (b) indicates this input power which gives minimum cycle-slip free stabilization power. The intersection of the dotted purple and blue lines is the estimated number of photons when the quantum channel is stabilized through the error signal from the stabilization channel with this extremely low power. The estimated photon flux of <0.01 photons/s is far below the background count rate of our detector. This number agrees well with the estimated photon flux of $8 \times 10^{-3}$ photons/s calculated from Eq. 1 and the measured Raman scattering spectrum on an optical spectrum analyzer (Fig. 4(a)). This shows that we achieved fiber link stabilization of 40 km MCF with negligibly low rate of Raman photons coupled into the quantum channel.

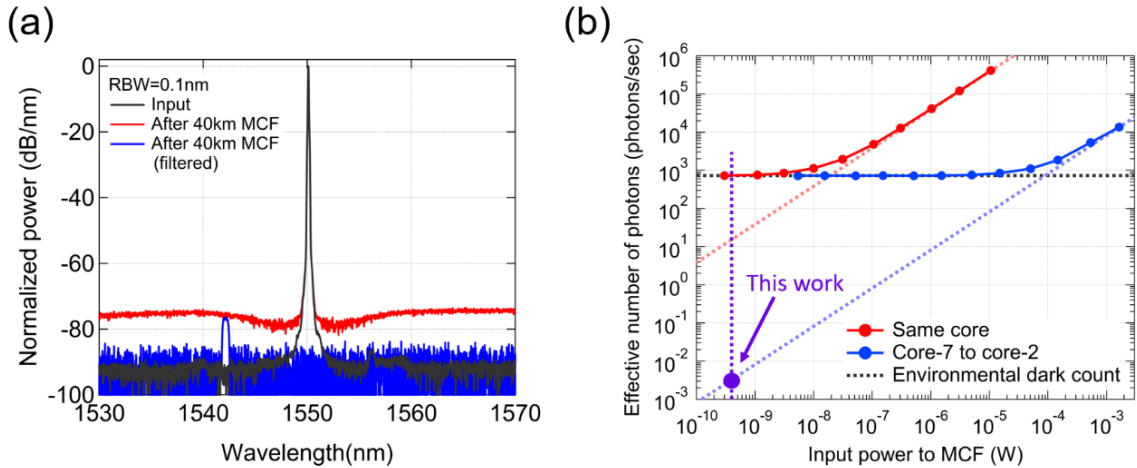

Figure 4. (a) Normalized input and spontaneous Raman scattering spectra at the output of core-7 after 40 km MCF. The spectra are measured using an optical spectrum analyzer. (b) Number of noise photons measured by the SNSPD when using a single core, a good approximation to SMF (red) or two cores (blue) of the MCF with different input power. Measurement floor is limited by black body radiation from the fiber to routing to the SNSPD. The extrapolated number of photons at the demonstrated power without cycle slips was below 0.01 photons/s (purple dot).

**Conclusion**

We investigated the relative core-to-core stability and Raman scattering crosstalk over 40 km of a 7-core fiber, towards co-existence of quantum and classical light for quantum networks. By separating the quantum and classical channels over different cores of the fiber and using WDM, we estimate a spurious photon rate on a 100% duty cycle quantum channel of <0.01 photons/s while maintaining cycle slip-free, sub-femtosecond relative timing jitter. This low spurious count was achieved with a combination of minimizing the launched power used for fiber path stabilization to 400 pW (with a corresponding power on the stabilization in-loop detector of only 300 fW) and 40 dB optical isolation between cores. Long-term time deviation measurements exhibited sub-femtosecond values for averaging times up to ~1000 s. This is the first demonstration of the co-existence over MCF with a complete study of spurious Raman photon coupling and low-power light stabilization. Our study strongly supports MCF as a promising solution for high-fidelity quantum networks with 100% duty cycle and sub-fs timing jitter.

**Acknowledgement**

We thank Jun Ye and JILA silicon cavity team for ultrastable reference light, and Kevin L. Silverman and Yifan Liu for helpful comments on the paper.